\documentclass[AMA,STIX1COL]{WileyNJD-v2}
\usepackage{moreverb}
\usepackage{textcomp}
\usepackage{booktabs}
\usepackage{here}
\usepackage{multirow}
\usepackage{url}
\usepackage{comment}
\usepackage{enumerate}
\usepackage{amssymb}
\usepackage{type1cm}
\usepackage{bm}
\usepackage{amsmath}

\def\BibTeX{{\rm B\kern-.05em{\sc i\kern-.025em b}\kern-.08em
    T\kern-.1667em\lower.7ex\hbox{E}\kern-.125emX}}

\newcommand{\bhline}[1]{\noalign{\hrule height #1}}

\newcommand{\ie}{\textit{i}.\textit{e}.~}

\newcommand{\myset}[1]{\{ #1 \}}

\articletype{}%

\received{}
\revised{}
\accepted{}

\begin{document}

\title{Topic Allocation Method on Edge Servers for Latency-sensitive Notification Service}

\author[1]{Tomoya Tanaka}

\author[2]{Tomio Kamada}

\author[3]{Chikara Ohta}

\authormark{T. Tanaka \textsc{et al}}

\address[1]{\orgdiv{School of Information Technologies}, \orgname{Tallinn University of Technology}, \orgaddress{\state{Tallinn}, \country{Estonia}}}

\address[2]{\orgdiv{Graduate School of System Informatics}, \orgname{Kobe University}, \orgaddress{\state{Kobe}, \country{Japan}}}

\address[3]{\orgdiv{Graduate School of Science, Technology and Innovation}, \orgname{Kobe University}, \orgaddress{\state{Kobe}, \country{Japan}}}

\corres{*Tomoya Tanaka, Ehitajate tee 5, 12616 Tallinn, Estonia \email{totana@ttu.ee}}

\presentaddress{Akadeemia Tee 7/2, 12611, Tallinn, Estonia}

\abstract[Abstract]{
The importance of real-time notification has been growing for social services and Intelligent Transporting System (ITS). As an advanced version of Pub/Sub systems, publish-process-subscribe systems, where published messages are spooled and processed on edge servers, have been proposed to achieve data-driven intelligent notifications. 
In this paper, we present a system that allows a topic to be managed on multiple edge servers so that messages are processed near the publishers, even when publishers are spread over a wide area. 
Duplicating messages on geographically distributed servers could enable immediate notification to neighboring subscribers.
However, the duplicated message spool may cause exhaustion of resources. 
We prepare a formal model of our publish-process-subscribe system and formulate the topic allocation as an optimization problem under the resource constraints of edge servers.
As the optimization problem is NP-hard, we propose heuristics leveraging the locality and the pub/sub relationships observed between clients to use the edge server resources efficiently. Our performance evaluation shows that our method reduces the delay to deliver notifications and the effectiveness of the strategy exploiting the relationships between clients.
}

\keywords{Real-time notification; multi-access edge computing; publish-process-subscribe model}


\maketitle

\section{Introduction}\label{sec:section1}
In recent years, the importance of real-time notification combined with the use of data has been growing for applications such as IoT applications and Intelligent Transporting System (ITS).\cite{mec_application_requirements} For example, real-time decision-making services proposed in ITS are expected to react immediately to changes in traffic conditions, analyze the current conditions, and provide optimal behavior for vehicles.\cite{its_2} Multi-access Edge Computing (MEC) and Pub/Sub messaging models have been exploited to generate notifications immediately, reflecting the changing spatio-temporal conditions.

MEC, proposed by the European Telecommunications Standards Institute (ETSI), enables ultra-low latency and high bandwidth by geographically distributing computation and storage resources to edge servers.\cite{mec_2014} MEC has already been applied to IoT, AR / VR, and traffic management.\cite{mec_application_requirements} On the other hand, Pub/Sub model can be adapted to the design of large-scale distributed systems such as MEC.\cite{pubsub_max} In particular, topic-based and content-based Pub/Sub model has been used in social interaction message notifications and event notifications among vehicles.\cite{pubsub_sns,pubsub_vehicles} 
A more sophisticated version of Pub/Sub called publish-process-subscribe enables data-driven notifications by spooling and analyzing messages.\cite{pub-pro-sub}

Many studies assume that one topic is managed by one server. We present a topic-based publish-process-subscribe system that allows one topic to be managed by multiple edge servers.
By managing one topic at multiple edge servers, publishers can send messages to the nearby edge server which manages the message's particular topic. The server generates the notifications for the messages and delivers them to subscribers.
With such a system, duplicating messages on geographically distributed servers could enable immediate notifications from publishers to neighboring subscribers. 


However, due to the limited resources on edge servers, duplicating messages on numerous edge servers could cause exhaustion of storage capacity. 
An efficient way of using computational resources consumed by Pub/Sub brokers, which perform analysis of the streaming messages in the publish-process-subscribe paradigm, has been presented.\cite{pub-pro-sub-app}
In contrast, we mainly focus on utilizing storage capacity on edge servers efficiently while enabling one topic to be managed on multiple edge servers so that the published messages are processed at nearby edge servers and the generated notifications are immediately delivered to neighboring subscribers.
Note that while our scheme increases the total storage capacity needed for duplicated message spools, it does not increase the total computational cost for message processing executed in a distributed manner.

We have proposed the prototype design of our publish-process-subscribe system and conducted a preliminary experiment to show the tradeoff between network latency and the required storage capacity.\cite{tomoyaComEX20}
In this previous work, we measured the path length from publishers to subscribers and the storage capacity used on each edge server depending on the number of edge servers managing one topic. We ignored the delay caused by concentrating clients on specific servers. 

In this paper, we propose a model that links the path from publishers to subscribers and estimates the consumed storage capacity and computational resources on each edge server.
Based on this model, we propose a method to allocate topics on edge servers in order to achieve real-time notifications in a publish-process-subscribe system under limited resources.
Currently, our scheme does not assume clients that change their location or message pattern dynamically.
This paper makes the following key contributions:
\begin{itemize}
\item  We propose a delay model of the notification delivery from a publisher to subscribers and formulate the topic allocation method as an optimization problem of the delay under the resource constraints on edge servers. 
\item As the formulated optimization problem is NP-hard, we propose a heuristic named RELOC, which allocates topics on edge servers so that notifications can be processed at neighboring edge servers immediately. We exploit locality and topic-derived relationships observed among publishers and subscribers to use storage capacity efficiently while maintaining proximity between clients and a message processor.
\end{itemize}
This paper is an extended version of previous work published in APNOMS 2020.\cite{tomoyaAPNOMS20}
We explain the related work and the motivational case in more details, enhance the model used to represent computational resources on edge servers, and improve the RELOC algorithm and its evaluation.

The remainder of this article is organized as follows: Section~\ref{sec:section2} explains related works to clarify the position of our research. The overview of our system is presented in Section~\ref{sec:section3} and 
Section~\ref{sec:section4} illustrates the motivation case.
In Section~\ref{sec:section5}, we formulate the notification delivery delay for the presented publish-process-subscribe system. In Section~\ref{sec:section6}, we elaborate on our topic allocation method RELOC. Simulation results are shown to evaluate the performance of our topic method in Section~\ref{sec:section7}.

\section{Related Work}
\label{sec:section2}
Topic-based Pub/Sub systems categorize messages using topics. The messages are not directly sent from senders (called \textit{publishers}) to receivers (\textit{subscribers}). Instead, a \textit{broker} prepared for each topic receives messages published to this topic and delivers them to the subscribers of this topic.
Many Pub/Sub systems assume there is a unique broker per topic and do not support distributed broker situations.
In cloud computing, scalable Pub/Sub systems are used to connect multiple servers with low latency.\cite{kafka,goodhope2012building,cloudPubSub}  Apache Kafka\cite{kafka}, which is developed by LinkedIn, can allocate multiple brokers for each topic and supports consumer groups. Each record published to a topic is delivered to one consumer in a group. This system focuses on scalability and fault-tolerance of the distributed system but pays little attention to the geographical locality.

Publish-process-subscribe systems are an evolution of Pub/Sub mechanisms where published messages are processed before disseminating them to subscribers. Publish-process-subscribe model based real-time communication is investigated in several recent studies.\cite{pub-pro-sub}
The message processing causes critical problems in achieving real-time notifications due to the processing load. Khare et al. present techniques to realize a scalable broker architecture that balances data publication and processing load for publish-process-subscribe systems operating at the edges, and ensures Quality-of-Service (QoS) on a per-topic basis.\cite{pub-pro-sub-app} 

Nevertheless, many researchers focusing on efficient edge server resource use assume that one topic is managed by one edge server. Occasionally, such one-topic-to-one-edge-server architecture does not guarantee the required QoS. For example, it could take longer to deliver messages from publishers to subscribers in applications where publishers and subscribers are spread over a wide area.\cite{wide} Besides, computational load and traffic load could be concentrated when numerous publishers and subscribers are allocated to one topic.
Managing a topic on multiple edge servers and brokers has the potential to achieve immediate message delivery from publisher to subscriber via their nearby edge server, as shown in Section~\ref{sec:section4}.
This one-topic-to-many-edge-server architecture could distribute computational load and traffic load. However, preparing message processors on multiple servers could cause storage capacity over-consumption.

Considering the limited resources of edge servers, many researches focus on proactively fetching content on edge servers to improve latency for clients to obtain content. The criteria to determine where and which content should be cached include: (1) content request probability\cite{pc_popu_size, cc_tl}, where the most popular content is cached on edge servers; or (2) client mobility\cite{pc_user_mob, pc_qlearning}, where content is cached on an edge server near the client who will request the content. In contrast, in publish-process-subscribe systems, the messages or data objects to be stored on edge servers is determined by topics and publishers/subscribers relationships.
In this research, we propose an efficient use of the limited storage capacity and other resources of edge servers with adequate allocation of topics to edge servers. To the best of our knowledge, topic-allocation-based messages management on edge servers in publish-process-subscribe paradigm has not been presented before.

Some distributed storage services are proposed to offer storage service for distributed edge servers.
FogStore\cite{fogStore17,fogStore18} is a distributed data store for Fog/edge computing.
It uses a distributed data store for cloud data centers as a base system,
and introduces location-conscious replica placement strategy using a context (location) based index.
It places some replicas on neighboring hosts to enable efficient quorum based queries,
and also allocates remote replicas to provide tolerance from geographically-correlated failures.
These works only considers the location context of data
and do not support publish/subscribe relationships between remote clients.
ElfStore\cite{elfstore} is an edge-local federated store over unreliable MEC hosts.
The storage namespace has a flat set of streams, and MEC hosts can add data blocks to the streams.
It features a federated indexing model using Bloom filters and differential replication scheme across edge servers, but it does not consider data processing or notifications to subscribers.
Nagato et al. propose a distributed data framework in a pub/sub paradigm.\cite{nagato}
It allows application developers to define proactive caching rules based on pub/sub relationships, but it does not take into account resource capacity or data processing.

\section{System Overview}
\label{sec:section3}

We consider a publish-process-subscribe system, where each edge server generates and transmits data-driven notifications near publishers so that immediate notifications are delivered to the nearby subscribers.

\begin{figure}
  \begin{center}
      \includegraphics[width=10cm]{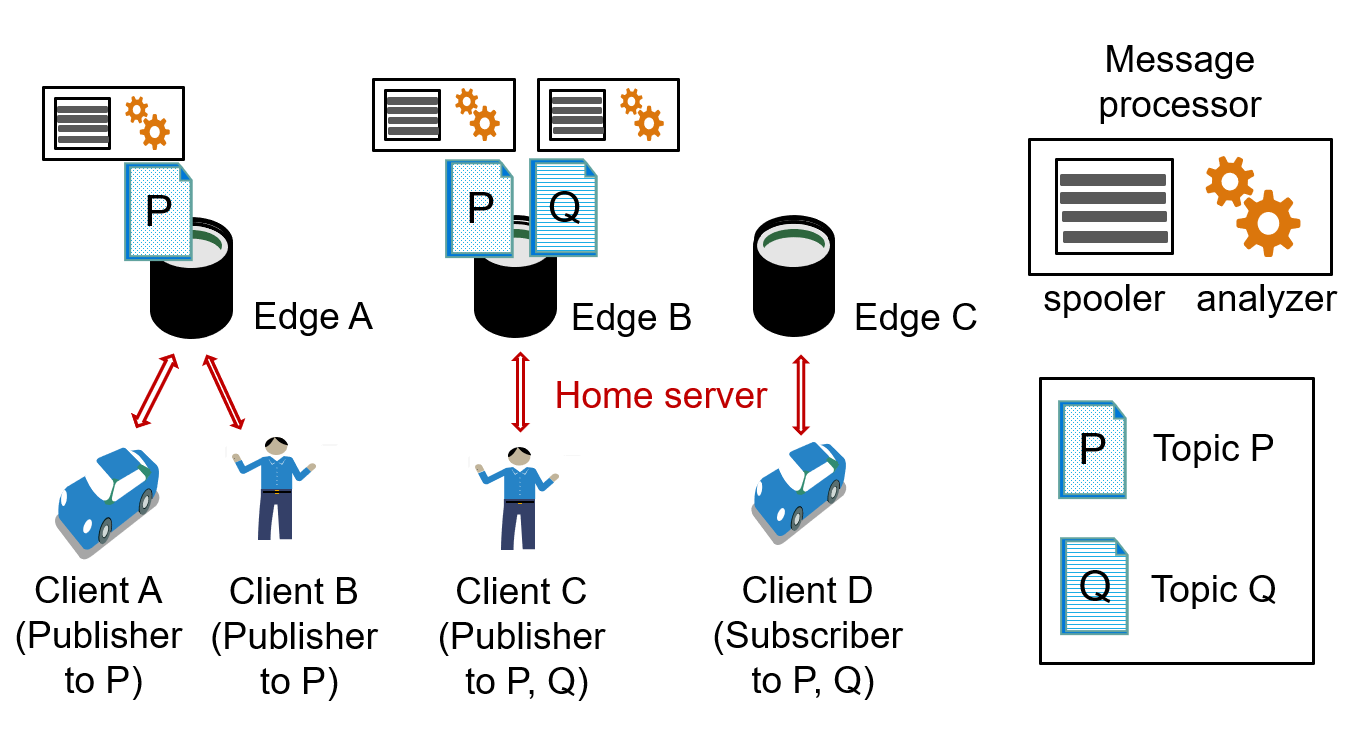}
    \caption{Overview of our publish-process-subscribe system}
    \label{fig:arch}
  \end{center}
\end{figure}

FIGURE~\ref{fig:arch} presents an overview of our system, where each publisher is assigned to a single edge server which manages all the topics the publisher registers. 
For example, Client C publishes messages to Topic P and Q, and  Edge B - the assigned server of Client B - manages both of these topics.
We call the server assigned to each client the ``home server.''
A message processor is prepared for each topic that an edge server manages.
It functions as a message spooler and an analyzer of the spooled messages for a topic.

\begin{figure}
  \begin{center}
      \includegraphics[width=10cm]{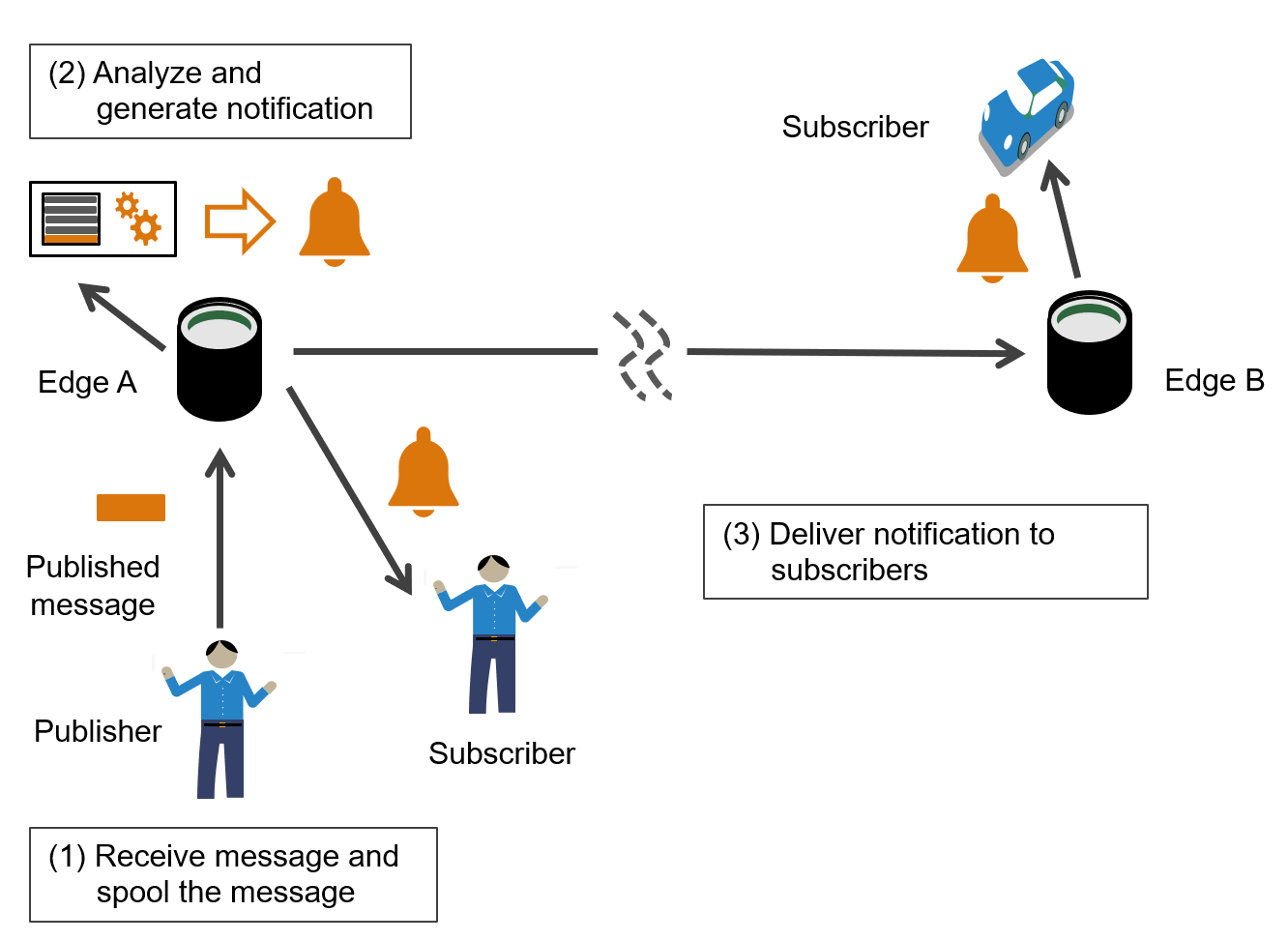}
    \caption{Notification delivery flow}
    \label{fig:flow}
  \end{center}
\end{figure}

On this topic management mechanism, the system generates notifications and delivers them to subscribers as shown in FIGURE~\ref{fig:flow}.
When an edge server receives a message from a publisher, it spools messages in the message processor prepared for the topic.
Next, the message processor analyzes the spooled messages and generates a notification.
Finally, the generated notification is delivered to the subscribers.
Note that, message processing is conducted only by the edge server which receives the message from the publisher. The generated notifications are either directly sent to the subscribers, or delivered via the subscribers' home server.

The home server also functions as a broker which receives messages from publishers and delivers the notifications produced by the message processors to the clients.
In FIGURE~\ref{fig:arch}, Edge B manages topics P and Q, that is, Edge B receives messages for topics P and Q and generate notifications for both of these topics.
The notifications are then transferred to clients A and C via their home servers.
In this case, Edge C has a single client which only subscribes to messages. As a consequence, it does not have any message processors.

\section{Motivational Case}
\label{sec:section4}

In this section, we explain the motivational case of our system and discuss the advantage of the proposed system.
Let us consider an application in which data is collected from pedestrians, automobiles and IoT sensors. Our system tries to generate beneficial information such as alerts using the collected data.

\begin{figure}
  \begin{center}
      \includegraphics[width=17cm]{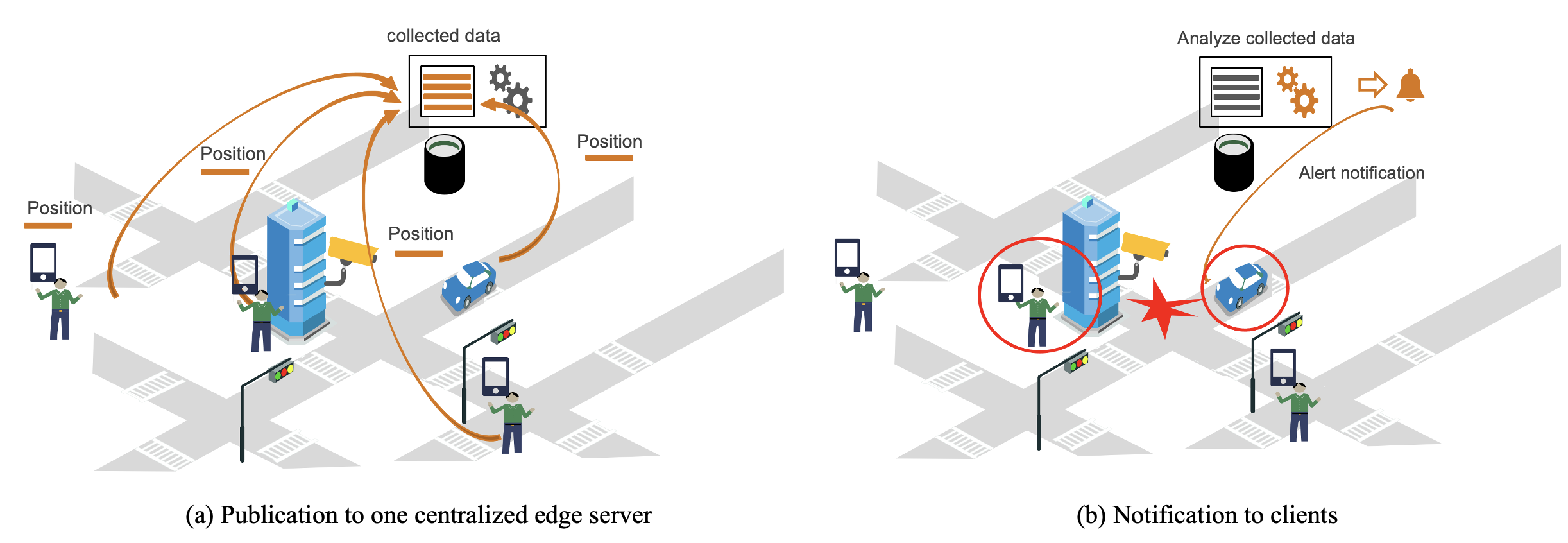}
    \caption{Message Publication and notification delivery with a centralized edge server}
    \label{fig:default}
  \end{center}
\end{figure}

The FIGURE~\ref{fig:default}a shows a situation where messages containing the positions of pedestrians and automobiles are collected to a single server. Based on the collected data, our publish-process-subscribe system generate a notification.
In the situation presented in FIGURE~\ref{fig:default}b, the pedestrian circled in red is suddenly leaving the building, and is in danger of being hit by the car circled in red. In such a situation, our system would generate an alert notification based on the positions of the car and the pedestrian, and deliver it to both the car and the pedestrian. After the automobile and the pedestrian receive the notification, they can both take appropriate actions to avoid an accident.

\begin{figure}
  \begin{center}
      \includegraphics[width=17cm]{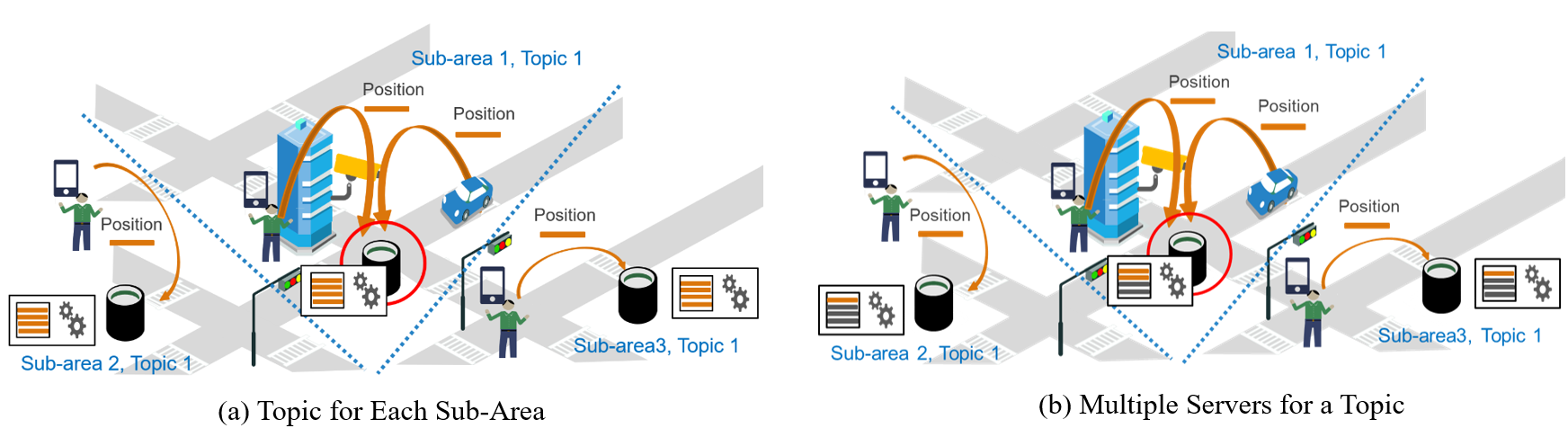}
    \caption{Message Publication and notification delivery with multiple edge server}
    \label{fig:multiple}
  \end{center}
\end{figure}

Next, let us consider a situation where multiple edge servers are deployed in the field.
In FIGURE~\ref{fig:multiple}, an edge server is placed at each intersection, for a total of three in this area.
More generally, the area is separated into three sub-areas and one topic is assigned for each sub-area.
In the case of FIGURE~\ref{fig:multiple}a, the message and notification delivery from publishers to subscribers in the same sub-area will be faster compared to the centralized case presented in FIGURE~\ref{fig:default}.
However, the system cannot generate alert notifications using messages in different sub-areas.
For instance, if automobiles runs with high speed, it is better to share information over a wider range. Moreover, if surveillance cameras can recognize pedestrians behaving erratically (either drunk or ignoring signals), this information should be shared within a wider area. 

To resolve this, we propose another topic allocation shown in FIGURE~\ref{fig:multiple}b, where one topic is assigned to all three of the edge servers in the area.
The automobiles and pedestrians publish messages to the closest edge servers, while the broker which receives the messages copies them to the other edge servers.
Such an allocation will triple the storage space required. However, messages and notifications delivery between publishers and subscribers in the same sub-area will be faster than the case presented in FIGURE~\ref{fig:default} and as fast as the case presented in FIGURE~\ref{fig:multiple}a.
Note that a server which receives a message from a publisher can utilize the spooled messages from other sub-areas to generate notifications.
Notifications delay between clients in different sub-areas does not improve, but it makes sense to give notifications of imminent danger between clients in closer positions higher priority.

Our objective in the following sections is to investigate topic allocation to achieve minimum delay for the message and notification delivery from a publisher to subscribers, taking into account constraints such as the available storage capacity on edge servers, and the number of connections between edge servers and clients. The topic allocations shown in FIGURE~\ref{fig:multiple}b could achieve lower latency for notification delivery, but it is too naive to use edge server resources efficiently.
Note that this example includes many mobile clients, even though our scheme does not currently care for the mobility of clients because of some reason described in Section~\ref{sec:section5:3}.
We will tackle these challenges in future work.

When the calculated minimum delay is lower than the application requirements, the application can operate their service as expected. Otherwise, they can obtain more storage capacity or network resource and reconfirm that their requirement is met with the best topic allocation.

\newcommand{\distCS}[2]{ d^{\texttt{H}}_{{#1},{#2}} }
\newcommand{\distSS}[2]{  d^{\texttt{S}}_{{#1},{#2}} }
\newcommand{\distCloud}{  d^{\texttt{C}} }
\newcommand{\rl}{  \theta }

\section{System model and problem formulation}
\label{sec:section5}

In this section, we introduce a formal model of our publish-process-subscribe system. 
Given an assignment of a home server to each client, the model estimates the delay in delivery of a message from the publisher via the nearby message processors to subscribers.
The optimal allocation of topics under the resource constraints of edge servers is formalized as an optimization problem of the total sum of the delay.

To treat the situation where topics assigned to an edge server needs more storage capacity or computation resource than the edge server has,  we introduce a delegation mechanism of message processing to the cloud server.
When the assigned server of a client does not have a message processor for a topic, the published messages are forwarded to the cloud server, incurring a delay in visiting the cloud server.



In the following sections, we show the influence of topic allocation on the delivery delay within our publish-process-subscribe system model followed by a formal definition of our optimization problem.


\subsection{System model}
We consider a field with $L$ edge servers denoted $\mathcal{S} = \{s_1, s_2, \cdots, s_L\}$. In the field, $M$ clients, denoted $\mathcal{C} = \{c_1, c_2, \cdots, c_M\}$, publish or receive messages. Each client $c_m \in \mathcal{C}$ has a home server $s_l \in \mathcal{S}$.

Messages are managed on a topic basis. Let $\mathcal{T} = \{t_1, t_2, \cdots, t_N\}$ denote the set of all topics that have been created for the system, and $\mathcal{T}_l \subseteq T$ denote the set of topics that are managed by the server $s_l$. 
Finally, let $\mathcal{C}_{n} \subseteq \mathcal{C}$ denote the set of clients who are publishers or subscribers of topic $t_n$ and let $\mathcal{C}^{\text{P}}_{n, l} \subseteq \mathcal{C}_n$ denote the set of publishers of topic $t_n$ whose home server is $s_l$.
A topic $t_n$ has its own message processor that uses a message spooler of size $u_n$.
The size of the message spoolers for each topic is predefined to hold enough entries needed for analysis to generate notifications.
If a message spooler reaches its capacity limit, it discards the oldest messages to store new ones.
We introduce two resource constraints of edge servers, namely storage capacity limit and computational resource limit. Let $A_l$, and $B_l$ denote the storage capacity limit, and the computational resource limit on server $s_l$ respectively. Each server $s_l$ cannot hold message spoolers greater than $A_l$, while each edge server cannot have computational load over $B_l$.

\begin{figure}
  \begin{center}
      \includegraphics[width=10cm]{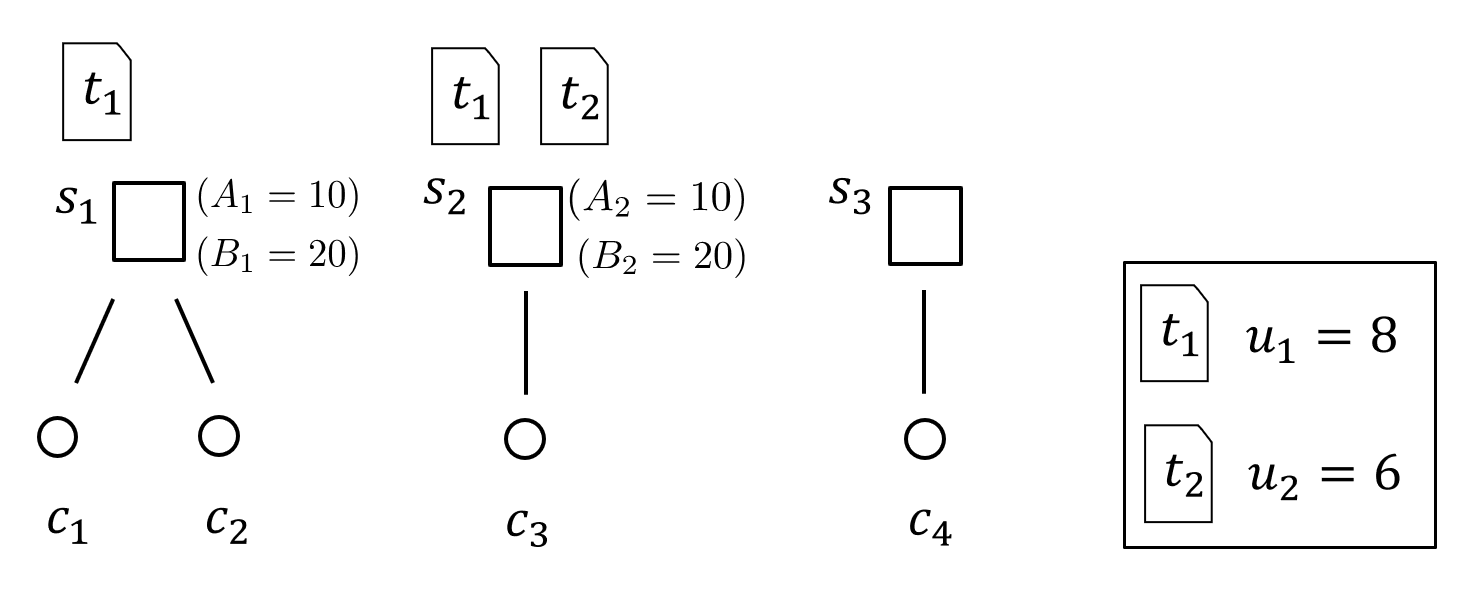}
    \caption{System representation}
    \label{fig:model}
  \end{center}
\end{figure}

Let us consider an example. FIGURE~\ref{fig:model} is a simplified representation of the topic allocation and home server assignment shown in FIGURE~\ref{fig:arch}. In this case, $\mathcal{S} = \{s_1, s_2\}$, while $T = \{t_1, t_2\}$. Moreover, $\mathcal{C} = \myset{c_1, c_2, c_3, c_4}$, where $c_1$ and $c_2$ are publishers of $t_1$; $c_3$ is a publisher of $t_1$ and $t_2$; $c_4$ is a subscriber of $t_1$ and $t_2$.
$\mathcal{C}_1 = \{c_1, c_2, ,c_3, c_4\}$ and $\mathcal{C}_2 = \{c_2, c_4\}$.
Now, the size of the message spooler for topic $t_1$ and $t_2$, namely $u_1$ and $u_2$, are $8$ and $6$, respectively.
$A_1$ and $A_2$ are configured as $10$, while $B_1$ and $B_2$ are configured as $20$.

The total number of spooled messages for all topics assigned to server $ s_l $ can be expressed as $\sum_{t_n \in \mathcal{T}_l} u_n$.
The computational load to process all the topics assigned to server $s_l$ is denoted as $v_l$. 
We assume that the computational load is proportional to the number of analyzed messages stored in the message spooler and the number of clients who publish messages to the topic.
$v_l$ is calculated by
\begin{equation}
  v_l = \sum_{t_n \in \mathcal{T}_l}|\mathcal{C}^{\text{P}}_{n, l}|  u_n.  
  \label{equ:v_n}
\end{equation}

When an edge server $s_l$ does not have enough capacity to spool messages for topics in $T_l$ or sufficient computational resource to analyze messages for topics in $T_l$, the edge server cannot host the message processors of some topics anymore. Instead, it delegates the message processing to the cloud server.
This situation happens when $\sum_{t_n \in \mathcal{T}_l} u_n > A_l$ or $v_l > B$.

In FIGURE~\ref{fig:model}, $\sum_{t_n \in \mathcal{T}_1} u_n = 8$ and $\sum_{t_n \in \mathcal{T}_2} u_n = u_1 + u_2 = 14$. On the other hand, $v_1 = u_1|\mathcal{C}^{\text{P}}_{1,1}| = 16$ and $v_2 = u_1|\mathcal{C}^{\text{P}}_{2,1}| + u_2|\mathcal{C}^{\text{P}}_{2,1}| = 14$.
As edge server $s_2$ reaches its storage capacity $A_2$, $s_2$ delegates the message processing of some topics to the cloud server.

We roughly estimate the delegation ratio $\rl_l$ of a server $s_l$  using the ratio of storage capacity and computation resource held by the server and the storage and computation required to process the messages locally. 
\begin{equation}
  \rl_l = \text{min}(\frac{A_l}{\sum_{t_n \in \mathcal{T}_l} u_n}, \frac{B_l}{v_l}, 1),
  \label{equ:ratio_l}
\end{equation}

This estimation may be inaccurate when considering resource fragmentation and/or the variations in frequency of message publication depending on the topics.
We adopt this rough estimation because the main purpose of this model is to find the optimal topic allocation.
Our model focuses on accurately estimating the delays depending on topic allocation in situations where delegation seldom occurs.



\subsection{Route for notification delivery}
In this section, we formalize the delay between a message being sent and the notification being delivered, which we denote $D$, considering the constraints of $A_l$ and $B_l$.
Let $\distCloud$ denote the network delay between any of the edge servers and the cloud server.
We use the same value for all the edge servers.
The delay in the path between a client $c_m$ and its home server $s_l$ is denoted by $\distCS{m}{l}$.
The delay in the path between servers $s_l$ and $s_{l^\prime}$ is denoted by $\distSS{l}{l^\prime}$.
These values are determined by the network distance between the locations of both servers.
In the performance evaluation in section~\ref{sec:section7}, we use the Euclidian distance instead of the network distance. 
We assume $\distCloud$ is much larger than $\distCS{m}{l}$ or $\distSS{l}{l^\prime}$.

\begin{figure}[t]
  \begin{center}
      \includegraphics[width=12cm]{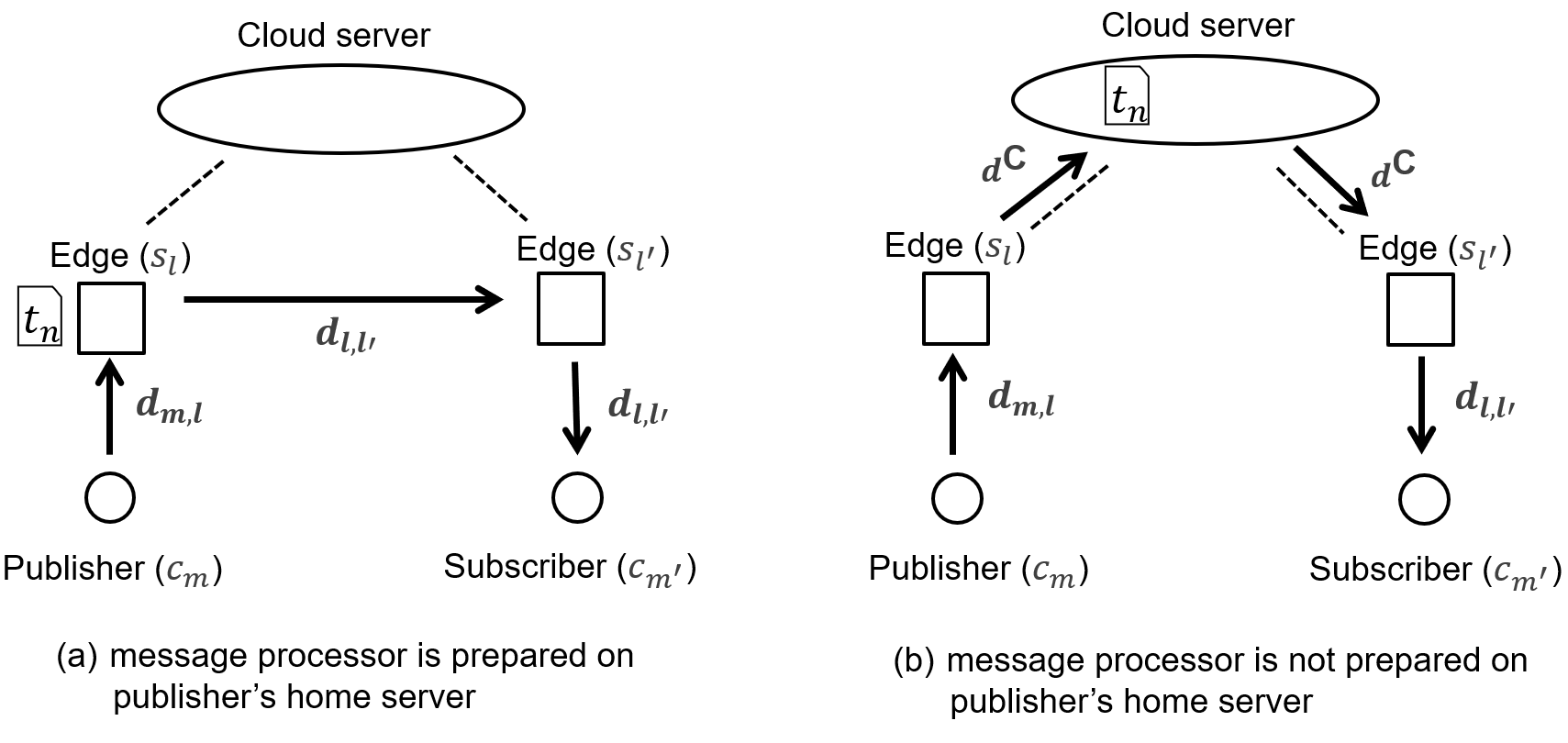}
    \caption{Route and process to deliver a notification to a subscriber}
    \label{fig:route}
  \end{center}
\end{figure}

FIGURE~\ref{fig:route} shows the route and process from publication of a message by a publisher $c_m$,
until the subscriber $c_{m^{\prime}}$ receives the corresponding notification.
(a) corresponds to the case where the message for the target topic is not delegated.
The home server of publisher $c_m$ is $s_l$, and the home server of subscriber $c_{m^{\prime}}$ is $s_{l^{\prime}}$.
(b) illustrates the case where the message processing is delegated to a cloud server.
In case of (a), $D$ equals to $\distCS{m}{l} + \distSS{l}{l^{\prime}} + \distCS{m^{\prime}}{l^{\prime}}$.
In case of (b), $D$ equals to $\distCS{m}{l} + 2\distCloud + \distCS{m^{\prime}}{l^{\prime}}$.

\subsection{Problem formulation}
\label{sec:section5:3}

The strategy of the optimization problem is denoted by a binary matrix $\bm{X} = \{x_{m,l} |c_m \in \mathcal{C}, s_l \in \mathcal{S}\}$. If $c_m$'s home server is $s_l$, $x_{m,l} = 1$, otherwise $x_{m,l} = 0$.
Our objective is to find the optimal $\bm{X}$ that minimizes the sum of the delay.


Keeping the ratio $\rl_l$ high and the value $\distCS{m}{l}$ small is incompatible.
When the publishers of the same topic are assigned to many edge servers, $\distCS{m}{l} $ will be small. However, 
the duplication of message processors increase the value of $\sum_{t_n \in \mathcal{T}_l} u_n$ and $v_l$ in Equation~\ref{equ:ratio_l}.
We have to take care not to concentrate topic allocations on any edge server to avoid capacity overflows.
Due to this trade-off, we try to find an optimal strategy under the given capacity constraints by formulating the problem mathematically.


Given $I$ message publications, we obtain an optimization problem which consists in adapting the topic assignment to edge servers to minimize the delay between message publication and notification delivery.
$\mathcal{P}_i$ denotes a set that only includes the publisher of the $i$-th message publication $|\mathcal{P}_i|=1$ and the $\mathcal{S}_i$ denotes the set of subscribes of $i$-th publication ($|\mathcal{S}_i|\ge1$).
We define the following optimization problem:

\begin{eqnarray}
  \mbox{min} && \sum_{i=1}^I D_i,
  \label{equ:objective}\\
  \mbox{s.t.} && x_{m,l} = \{0,1\}, \forall c_m \in \mathcal{C}, s_l \in \mathcal{S},
  \label{equ:const_1}\\
       && \sum_{l=1}^L x_{m,l} = 1, c_m \in \mathcal{C},
  \label{equ:const_2}
\end{eqnarray}
where
\begin{equation}
  D_i = \frac{1}{|S_i|}\sum_{c_m \in \mathcal{P}_i}\sum_{c_{m^{\prime}} \in \mathcal{S}_i} \Biggl[\sum_{l=1}^L \distCS{m}{l} x_{m,l}
  + \biggl[\sum_{l=1}^L 2\distCloud (1- \rl_l) x_{m, l} 
  +\sum_{l=1}^L \sum_{l^{\prime}=1}^L \rl_l \distSS{l}{l^{\prime}} x_{m, l}x_{m^{\prime} l^{\prime}}\biggr]
  + \sum_{l^{\prime}=1}^L \distCS{m^{\prime}}{l^{\prime}}x_{m^{\prime} l^{\prime}}\Biggr].
  \label{equ:d_i}
\end{equation}
In Equation~\ref{equ:d_i}, only when $\rl_l = 1$ (which means the edge server $s_l$ has enough capacity and computational resource to hold all the message processors of the topics it manages) does $1- \rl_l$ becomes $0$, otherwise $2\distCloud$ is added under the probability $1- \rl_l $. Similarly, $\distSS{l}{l^{\prime}}$ is added under the probability $\rl_l$. The optimization problem is NP-Hard because it is a kind of set partitioning problem.\cite{set_partition} Therefore, we develop heuristics based on an analysis of the formulated optimization problem.

Our system assumes that we know pub/sub communication patterns before solving the optimal assignments of home servers. This assumption may be satisfied if the targets are unrelocatable IoT sensor devices that periodically send signals.
On the other hand, if you want to apply this scheme to mobile agents, you may need to solve optimization problems periodically based on recent communication patterns.
Moreover, we have to take care of relocating message processors as publishers move.
We will tackle these challenges in future work.

\section{RELOC algorithm}
\label{sec:section6}
We develop a heuristic named RELOC (RElation and LOcality conscious Cooperative client assignment), which determines a topic-allocation strategy $\bm{X}$ which reduces the delay between message submission and notification delivery $D_i$, resolving the trade-off between $\rl_l$ and $d^{\text{H}}_{m,l}$. RELOC exploits notable features of publishers and subscribers, namely locality and topic-derived relationships. 

Initially, RELOC divides the field into $K$ clusters based on the position of edge servers using the $K$-means clustering algorithm. A client is only assigned to a server which belongs to the closest cluster so that the delay between the client and its home server $d^{\text{H}}_{m,l}$ is reduced.
The number of clusters $K$ is given in advance according to the strength of locality in client distribution. For example, in an application where publishers and subscribers are close together, we set $K$ larger to more finely divide the field. Adequate $K$ values are investigated in Section~\ref{sec:section7}.

Secondly, RELOC assigns publishers who have strong relationships to the same edge server. This helps reducing storage capacity consumption by using the same message processor to handle the topic to which these multiple publishers submit messages to. Initially, we prepare a list of publishers and subscribers for each topic $t_n$ (\ie $\mathcal{C}_1, \mathcal{C}_2, \cdots, \mathcal{C}_N$) from the given pub/sub communication pattern. Next, RELOC makes the matrix $\bm{Z} = \{ z_{m,m^{\prime}} |c_m, c_{m^{\prime}} \in \mathcal{C} \}$, where $z_{m,m^{\prime}}$ represents the number of times that both clients $c_m$ and $c_{m^{\prime}}$ are publishers/subscribers to the same topic (\ie $c_m \in \mathcal{C}_n \cap c_{m^{\prime}} \in \mathcal{C}_n$).
$z_{m,m^{\prime}}$ is the number of topics $t_n$ such that $c_m \in \mathcal{C}_n \cap c_{m^{\prime}} \in \mathcal{C}_n$.
After that, RELOC obtains the set of clients who have the strongest relationship with each other by adapting a matrix factorization technique\cite{matrix_factorization} to $\bm{Z}$. 

Note that publishers located in different clusters are not assigned to the same server even if they have a strong relationship.
Thus, RELOC can keep the distance between publishers and their home servers small within a cluster, while reducing the storage capacity consumption on the edge servers by assigning publishers who have strong relationships to the same edge server.

Finally, RELOC assign home servers to clients in order, taking into account the current resource usage of each edge server which is represented by $\rl_l$. If a resource on the client $c_m$'s home server $s_l$ is exhausted (\ie $\rl_l < 1$) as a result of placing a new message processor for topic $t_n$ on that server, a new home server  $s_{l^{\prime}}$ is assigned to client $c_m$. The topics that the client $c_m$ publishes to are copied to this new home server. RELOC then reassigns all the publishers to this topic $t_n$ from edge server $s_l$ to this new edge server $s_{l^{\prime}}$.
If there are no longer any clients who publish to a topic $t_n$ managed by server $s_l$ (\ie $\mathcal{C}^{\text{P}}_{n, l} = \phi$), RELOC removes this topic from this server.

The overall operation of RELOC is presented in Algorithm~\ref{alg:reloc}. The RELOC assigns home server to all clients $c_m \in \mathcal{C}$, where $\mathcal{C} = \mathcal{C}_1 \cup \mathcal{C}_2 \cup \cdots \cup \mathcal{C}_N$.

\begin{algorithm}
\caption{RELOC:home server assignment to clients}
\label{alg:reloc}
\begin{algorithmic}[1]
\Require The set of all topics $\mathcal{T}$; the size of message spooler of each topic $u_n$; storage capacity limit $A_l$; computational load limit $B_l$; The sets of clients who publish or subscribe to topic $t_1$, $t_2$, $\cdots$; $t_N$ (\ie $\mathcal{C}_1$, $\mathcal{C}_2$, $\cdots$, $\mathcal{C}_N$); the number of clusters $K$; the number of extracted clients who have the strong relationship $M^{\prime}$
\Ensure home server assignment strategy $\bm{X}$

\Function{Cooperation}{$c_m$, $s_{l^{\prime}}$, $s_l$, $T^{\prime}$}
  \For{$t_n$ \textbf{ in } $T^{\prime}$}
    \If{$t_n \textbf{ not in } T_l$}
      \State Add $t_n$ to $\mathcal{T}_l$
    \EndIf
    \State Add $c_m$ to $\mathcal{C}^{\text{P}}_{n, l}$ and update $\rl_{l}$ using the updated $\mathcal{T}_l$, $\mathcal{C}^{\text{P}}_{n, l}$
    \State Remove $c_m$ from $\mathcal{C}^{\text{P}}_{n, l^{\prime}}$ and update $\rl_{l^{\prime}}$ using the updated $\mathcal{C}^{\text{P}}_{n, l^{\prime}}$
  \EndFor
  \For{$t_n$ \textbf{ in } $T^{\prime}$}
    \If{$|\mathcal{C}_{n, l^{\prime}}| = 0$}
      \State Remove $t_n$ from $T_{l^{\prime}}$ and update $\rl_{l^{\prime}}$ using the updated $\mathcal{T}_{l^{\prime}}$
    \EndIf
  \EndFor
\EndFunction
\\
\State Initialize $\bm{X}$ by $\bm{X} = $ \textbf{0}
\State Divide field to $K$ clusters.
\State Makes a matrix $\bm{Z}$ based on $\mathcal{C}_1$, $\mathcal{C}_2$, $\cdots$, $\mathcal{C}_N$
\State Adapt matrix factorization to $\bm{Z}$ and obtain the set of clients who have the strongest relationship with each other
\For{$t_n \textbf{ in } \mathcal{T}$}
  \For{$c_m$ \textbf{ in } $\mathcal{C}_n$}
    \State $\mathcal{C}^{\prime}$ $\leftarrow$ $M^{\prime}$ clients who have the strongest relation with $c_m$
    \State $s_l$ : the home server of $c_m$  $\leftarrow$ the server that satisfies $\rl_l = 1$ and belongs to the same cluster as $c_m$ (initial value)
    \For{$c_{m^{\prime}}$ \textbf{ in } $\mathcal{C}^{\prime}$}
      \If{$s_{l^{\prime}}$ (home server of $c_{m^{\prime}}$)  and client $c_m$ belongs to the same cluster}
        \If{$\rl_{l^{\prime}} = 1$}
          \State $s_l$ $\leftarrow$ $s_{l^{\prime}}$
          \State \textbf{break}
        \EndIf
        \EndIf
     \EndFor
     \State $x_{m, l} = 1$
     \If{$c_m$ is a publisher of topic $t_n$}
     \State Add $t_n$ to $\mathcal{T}_l$ and $c_m$ to $\mathcal{C}^{\text{P}}_{n, l}$
     \State Calculate $\rl_l$ using the updated $\mathcal{T}_l$, $\mathcal{C}^{\text{P}}_{n, l}$
     \If{$\rl_l < 1$}
       \State $s_{l^\prime}$: previous home server $ \leftarrow s_{l}$
       \State $s_l$ $\leftarrow$ an edge server with minimum resource usage which belongs to the same cluster as $c_m$ 
       \State $x_{m, l^{\prime}} = 0$, $x_{m, l} = 1$
       \State $\mathcal{T}^{\prime}$ $\leftarrow$ the set of topics to which client $c_m$ publishes messages
       \State \Call{Cooperation}{$c_m$, $s_{l^{\prime}}$, $s_l$, $\mathcal{T}^{\prime}$}
    \EndIf
    \EndIf
  \EndFor
\EndFor
\end{algorithmic}
\end{algorithm}

\begin{enumerate}
\renewcommand{\labelenumii}{(\arabic{enumi})}
\item \textit{locality conscious:} In Line 17, the field with multiple edge servers is divided into $K$ clusters using the $K$-means clustering algorithm.
\item \textit{relation conscious:} In Line 19, the set of clients who have the strongest relationship are obtained.
  In Line 22, $M^{\prime}$ clients who have the strongest relationship with $c_m$ are stored in $\mathcal{C}^{\prime}$. A home server $s_l$ is temporarily selected from servers which belong to the same cluster as $c_m$ and satisfies $\rl_l = 1$. If extracted clients include a client who belongs to the same cluster as $c_m$, $s_l$ is updated to index of the extracted user's home server. In Line 32, RELOC update $\bm{X}$ and update server states $\mathcal{T}_l$, $\mathcal{C}^{\text{P}}_{n, l}$ as well as $\rl_l$.
\item \textit{cooperation:} If a resource on the home server $s_l$ of client $c_m$ is exhausted (\ie $\rl_l < 1$) after updating $\rl_l$, a new home server is assigned to client $c_m$ in Line 38. Topics that the client $c_m$ publishes (\ie $T^{\prime}$) are copied to the new home server in Line 4. In Line 11, if all $ c_m \in \mathcal{C}^{\text{P}}_{n, l}$ have been assigned to a new home server (\ie $\mathcal{C}^{\text{P}}_{n, l} = \phi$), then RELOC removes the topic and update $a_{l^{\prime}}$ and $b_{l^{\prime}}$.

\end{enumerate}

The time complexity of RELOC is $O(|T| \cdot \text{max}(|\mathcal{C}_n|) \cdot \text{max}(|T^{\prime}|) \cdot |M^{\prime}|)$.
The variation of the complexity is shown in  Section~\ref{sec:section7}.

\section{Simulation}
\label{sec:section7}
In this section, we observe topic allocation conducted by RELOC algorithm, and evaluate the notification delivery delay given by the home server assignment conducted by RELOC. 

We obtain a list of topics and publishers/subscribers for each topic from a real-world social network dataset (\textit{user\_sns.txt}) provided by Tencent Inc.\cite{dataset} 
We group \texttt{Followee-userid} by \texttt{Follower-userid} and make the arranged dataset, where each line includes a follower and her/his followees as shown in FIGURE~\ref{fig:arranged_data}. 
We extract $N$ lines from the arranged dataset and assign a topic to each line. Besides, we regard a follower and her/his followees in each line as publishers and subscribers of the assigned topic as shown in FIGURE~\ref{fig:pubsublist}. 
In FIGURE~\ref{fig:pubsublist}, the set of publishers and subscribers of topic $t_1$ is $\{1000336, 1001726, 1006631, 1010402, \cdots \}$, the set publishers and subscribers of topic $t_2$ is $\{1000351, 1010402, 1029746, 1029906, \cdots \}$, and so on (\ie $\mathcal{C}_1 = \{1000336, 1001726, 1006631, 1010402, \cdots \}, \mathcal{C}_2 = \{1000351, 1010402, 1029746, 1029906, \cdots \}, \cdots, \mathcal{C}_N = \{1001787, 1004738, 1004398, 1001726, \cdots\}$).

From the obtained list of topics and publishers/subscribers for each topic, the set of topics $\mathcal{T}$ can be expressed as $\mathcal{T} = \{t_1, t_2, \cdots, t_N\}$, while the set of clients $\mathcal{C}$ can be expressed as $\mathcal{C} = C_1 \cup C_2 \cup \cdots \cup C_N$. 

\begin{figure}
\begin{center}
\begin{minipage}{0.6\textwidth}
    \lstset{%
      frame=tb,
    }
    \begin{lstlisting}
      follower, followees
      1000336, [1001726, 1006631, 1010402, ...]
      1000351, [1010402, 1029746, 1029906, ...]
      1001858, [1004622, 1006355, 1006824, ...]
      ...
      1001787, 1004738, 1004398, 1001726, ...
    \end{lstlisting}

\end{minipage}
\caption{Arranged dataset of \textit{user\_sns.txt}}
\label{fig:arranged_data}
\end{center}
\end{figure}

\begin{figure}
\begin{center}
\begin{minipage}{0.6\textwidth}
    \lstset{%
      frame=tb,
    }
    \begin{lstlisting}
      topic, publishers/subscribers of each topic 
      t_1, 1000336, 1001726, 1006631, 1010402, ...
      t_2, 1000351, 1010402, 1029746, 1029906, ...
      t_3, 1001858, 1004622, 1006355, 1006824, ...
      ...
      t_N 1001787, 1004738, 1004398, 1001726, ...
    \end{lstlisting}

\end{minipage}
\caption{Publishers and subscribers for each topic}
\label{fig:pubsublist}
\end{center}
\end{figure}

\begin{figure}
\begin{minipage}{0.48\textwidth}
  \begin{center}
      \includegraphics[width=8cm]{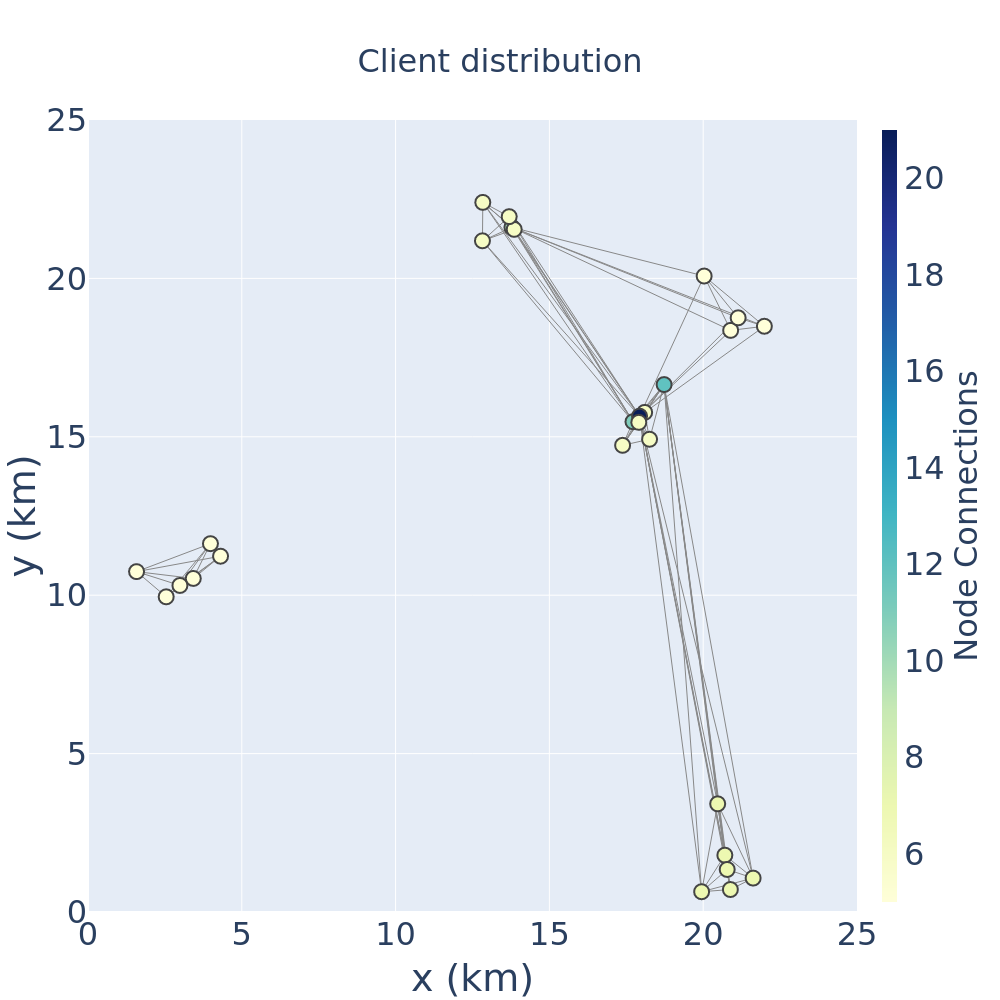}
  \end{center}
\end{minipage}
\hspace{0.3cm}
\begin{minipage}{0.48\textwidth}
  \begin{center}
      \includegraphics[width=8cm]{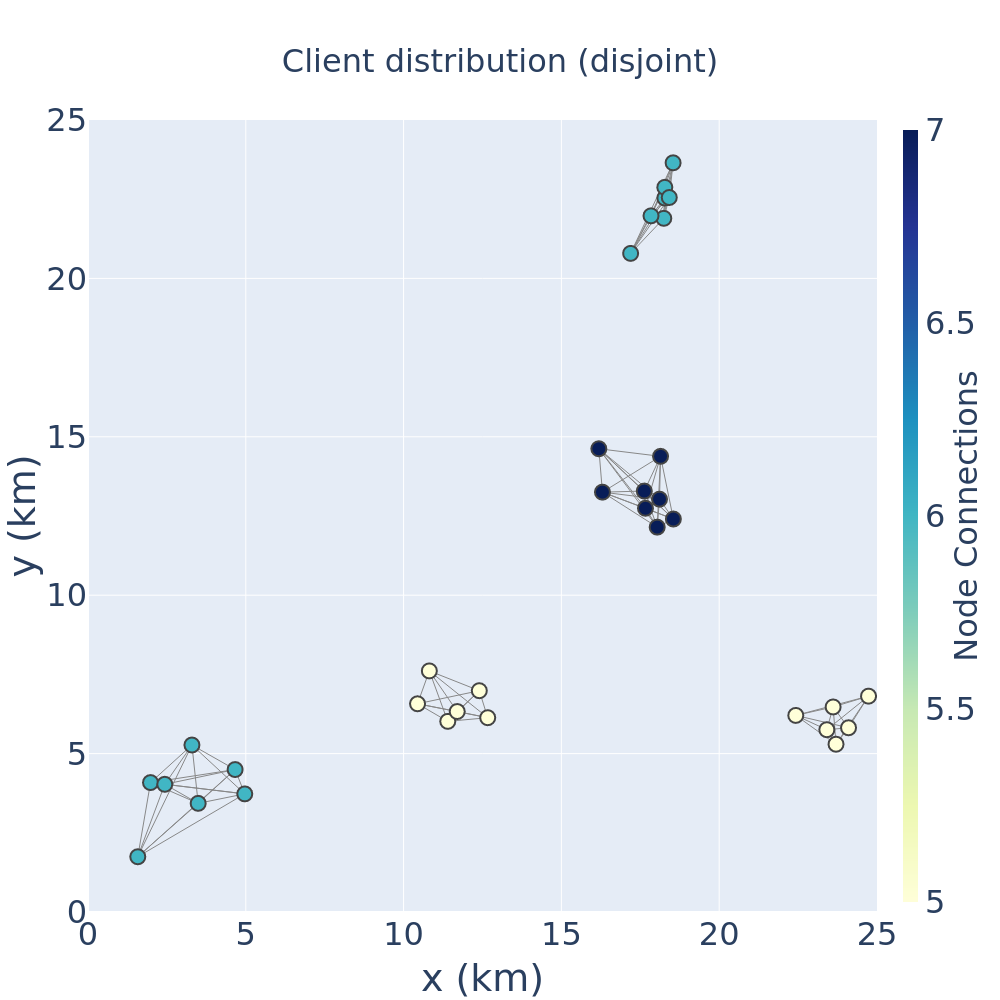}
  \end{center}
\end{minipage}
\caption{Examples of client distribution ($N = 5, L = 16, \sigma = 1$)}
    \label{fig:dist}
\end{figure}

We distribute the clients $c_m \in \mathcal{C}$ to a $25$~km square field, where publishers/subscriber of the same topic (\ie $c_m \in C_n$) follows a Gaussian distribution with a standard deviation of $\sigma$~km centered on a randomly chosen point in the field. Also, $L$ edge servers are placed in this field. FIGURE~\ref{fig:dist} shows examples of client distribution, where points represents clients. Two clients connected by an edge are publishers/subscribers of the same topic. The left figure shows the case that $C_n \cap \mathcal{C}_{n^{\prime}} \neq \phi$ for some $t_n, t_{n^{\prime}} \in \mathcal{T}$, while the right figure shows the case that $\mathcal{C}_n \cap \mathcal{C}_{n^{\prime}} = \phi$ for all $t_n, t_{n^{\prime}} \in \mathcal{T}$. 

\subsection{Preliminary evaluation}
As a preliminary evaluation, we observe a home server assignment determined by RELOC in a simple case. 
RELOC determines home servers of clients by Algorithm~\ref{alg:reloc} based on $T$, \{$\mathcal{C}_1$, $\mathcal{C}_2$, $\cdots$, $\mathcal{C}_N$\}, the size of message spooler of each topic $u_n$, storage capacity limit $A_l$, computational load limit $B_l$, the number of clusters $K$ and the number of extracted clients who have the strong relationship $M^{\prime}$. 

For sake of simplicity, let $A_1 = A_2 = \cdots = A_L = A$ and $B_1 = B_2 = \cdots = B_L= B$, in the following sections. 
Besides, all $c_m \in C$ can take a role of publishers and subscribers. This means that all clients publishes messages to their home server while receiving notification from their home server. On top of that, let us represent $u_n$ by $u_n = p|C_n|$, which means that the message processor of topic $t_n$ can spool $p$ messages published by a client on average. For example, if $|C_n| = 7$ and $p = 2$, the message processor of topic $t_n$ can hold $2|C_n| = 14$ messages. 

\begin{figure}[t]
  \begin{center}
      \includegraphics[width=17cm]{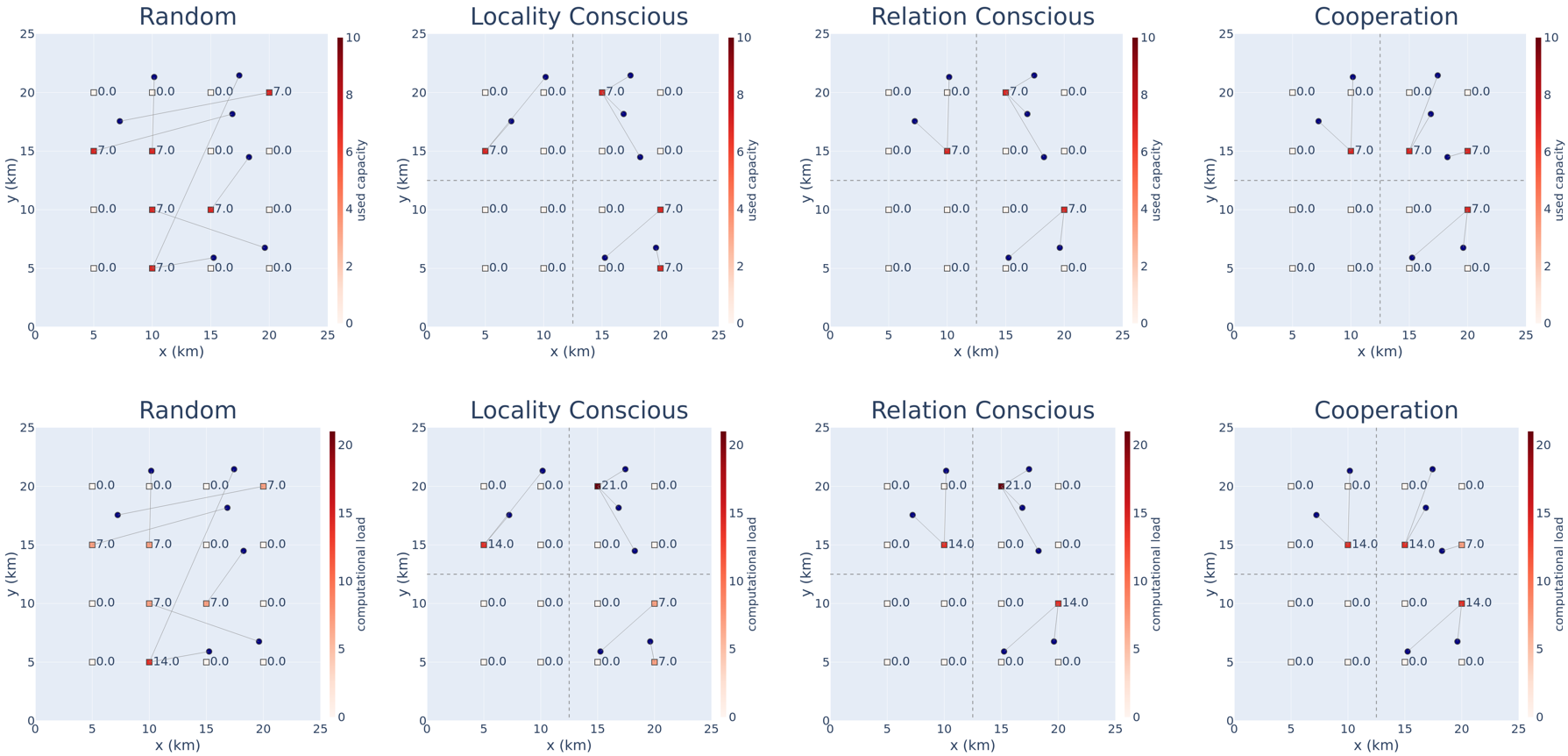}
    \caption{A home server assignments and the used capacity, computational load of edge servers after each step of RELOC}
    \label{fig:reloc_flow}
  \end{center}
\end{figure}

FIGURE~\ref{fig:reloc_flow} shows the result of home server assignment to clients by each step of RELOC, namely \textit{locality conscious}, \textit{relation conscious}, and \textit{cooperation}. The values of constants given in this simulation is shown in TABLE~\ref{tab:para_preliminary}.

\begin{table}[t]
 \caption{Values of constants (preliminary evaluation)}
 \label{tab:para_preliminary}
 \centering
 \begin{tabular}{c|c||c|c}
   \bhline{0.7pt}
   \textbf{Parameter} & \textbf{Value} & \textbf{Parameter} & \textbf{value}\\
   \hline
   $N$ & $1$ & $L$ & $16$\\
   $A$ & $7$ & $\sigma$ & $10$\\
   $B$ & $20$ & $p$ & $1$\\
   $K$ & $4$ & $M^{\prime}$ & $100$\\
   \bhline{0.7pt}
 \end{tabular}
\end{table}

\subsection{Performance Evaluation}
We compare the performance of RELOC with 3 assignment methods, namely Random Assignment (RA), the Nearest Server assignment (NS), and One-Topic-One-Server-assignment (OTOS). RA assigns home servers to clients randomly, while NS assigns the closest server to client. OTOS allocates one topic to one server, whereas RELOC allocates one topic to multiple servers. Note that OTOS does not work when a client is a publisher/subscriber of multiple topic such as a case shown in the right figure of FIGURE~\ref{fig:dist}. Therefore, we apply \textit{cooperation} mechanism to OTOS to prevent concentration of resource consumption.

FIGURE~\ref{fig:reloc_flow} shows the result of home server assignment to clients given by RA, NS, OTOS, RELOC. The values of constants are same as the values shown in TABLE~\ref{tab:para_preliminary} except for $A$ and $B$. In this method comparison, $A$ and $B$ are configured as $A=10$, $B=50$.

\begin{figure}[t]
  \begin{center}
      \includegraphics[width=17cm]{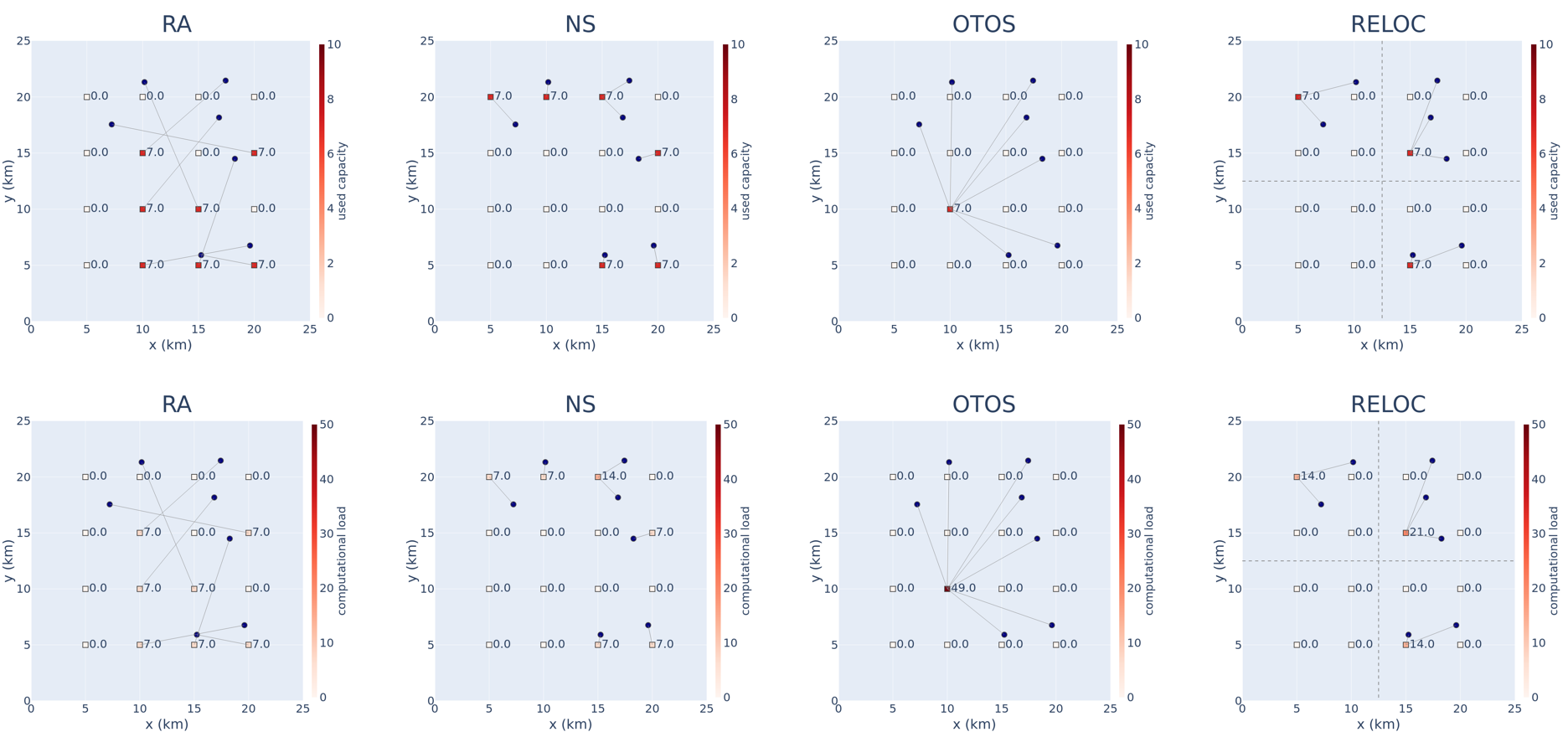}
    \caption{A home server assignment and the used capacity, computational load of edge servers given by RA, NS, OTOS, RELOC}
    \label{fig:comp_methods}
  \end{center}
\end{figure}

The simulation is conducted by the following steps. First, RA, NS, OTOS, and RELOC determines home server assignments based on $T$, \{$\mathcal{C}_1$, $\mathcal{C}_2$, $\cdots$, $\mathcal{C}_N$\}, $u_n$, $A$, $B$, $K$ and $M^{\prime}$. By this step, the resource usage of each edge server $\rl_l$ is calculated. The values of constants given in this simulation is shown in TABLE~\ref{tab:para_performance}. Values of $\sigma$, $p$, and $K$ are given in the simulation. 

\begin{figure}
\begin{minipage}{0.48\textwidth}
  \begin{center}
      \includegraphics[width=8cm]{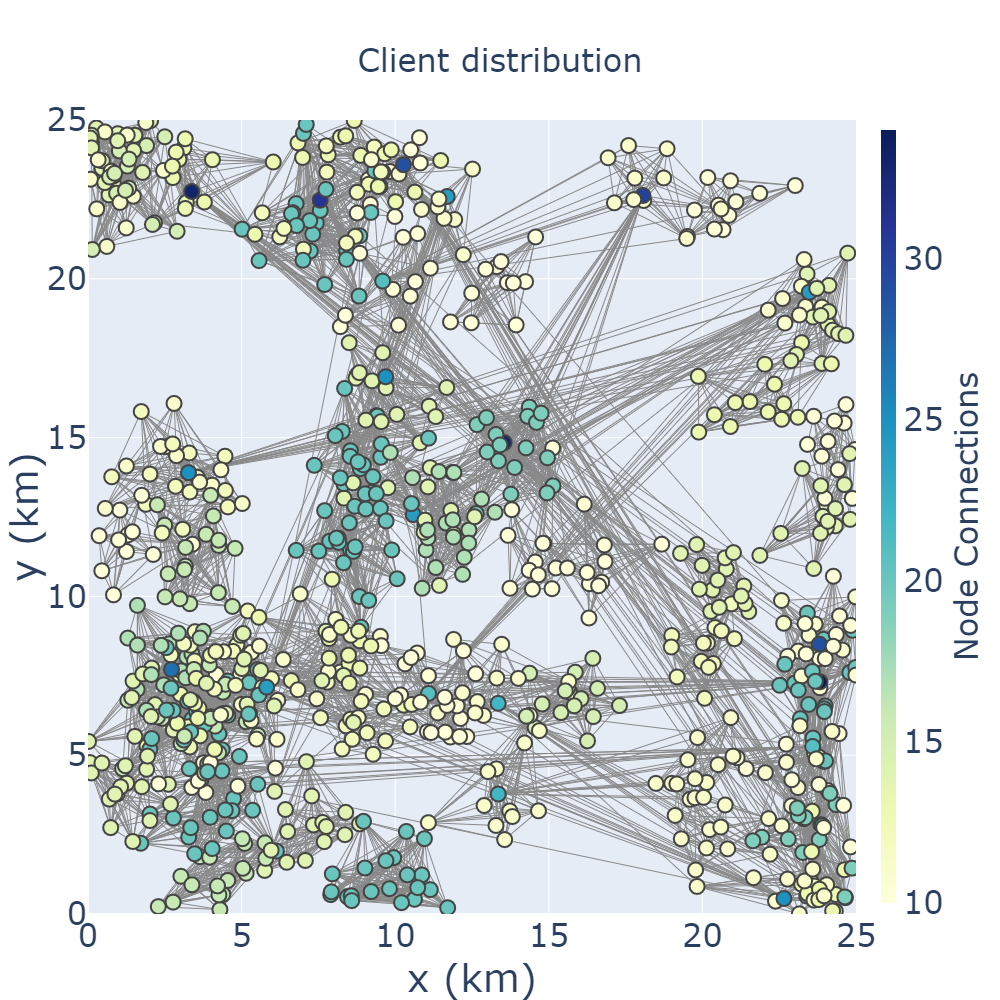}
  \end{center}
\end{minipage}
\begin{minipage}{0.58\textwidth}
  \begin{center}
      \includegraphics[width=8cm]{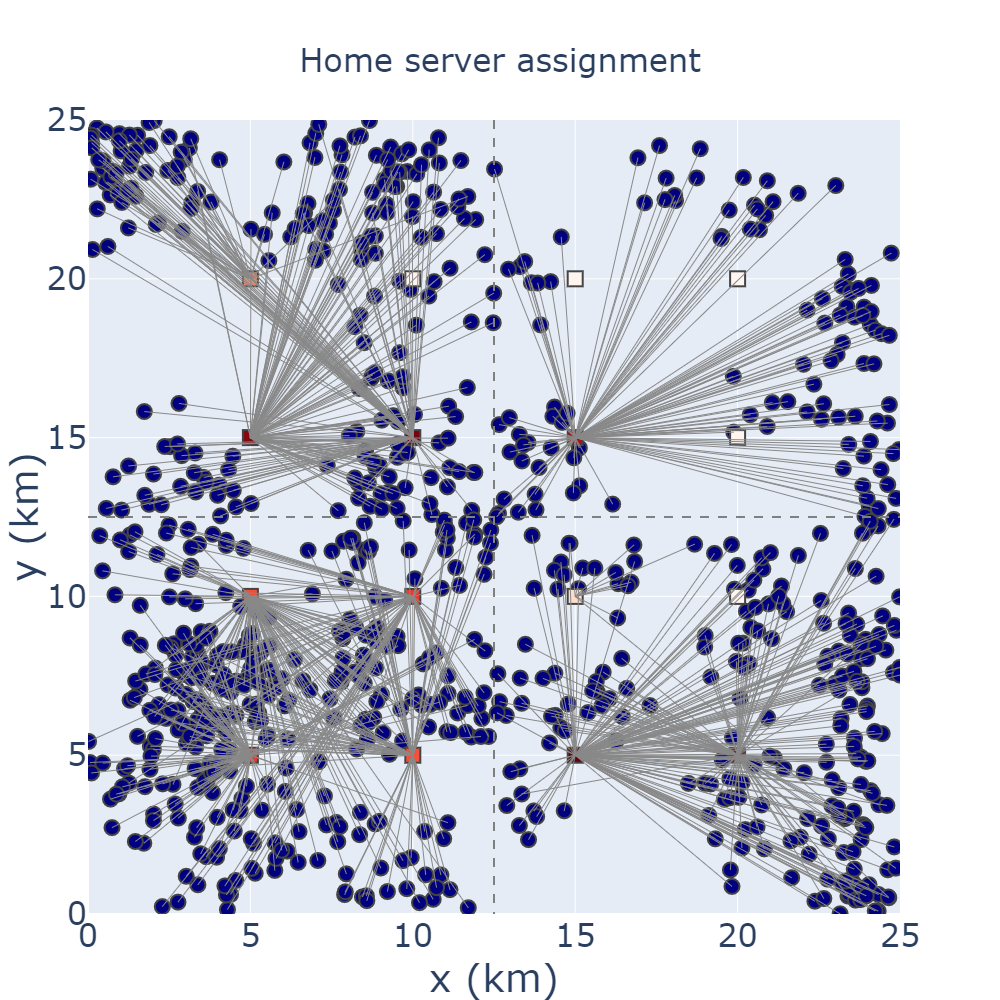}
  \end{center}
\end{minipage}
\caption{client distribution and home server assignment by RELOC ($N = 5, L = 16, \sigma = 1$)}
    \label{fig:distribution_performance}
\end{figure}

\begin{figure}[t]
  \begin{center}
      \includegraphics[width=8cm]{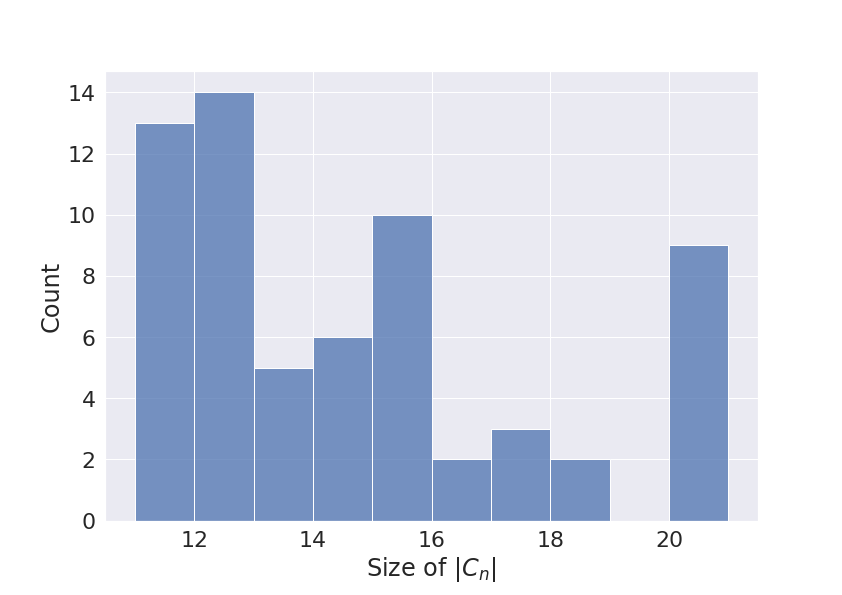}
    \caption{Distribution of the number of publishers/subscribers for each topic}
    \label{fig:hist_performance}
  \end{center}
\end{figure}

A client distribution and a home server assignment by RELOC when $\sigma = 1$, $p=2$, $K=4$ is shown in FIGURE~\ref{fig:distribution_performance}, where a client could be a publisher/subscriber of multiple topic (but less than $5$). The distribution of number of publishers/subscribers for each topic (\ie $|\mathcal{C}_1|, |\mathcal{C}_2|, \cdots |\mathcal{C}_N|$) is shown in FIGURE~\ref{fig:hist_performance}. Here, the values to calculate time complexity of RELOC, namely $|T| ,\text{max}(|\mathcal{C}_n|), \text{max}(|T^{\prime}|), |M^{\prime}|$ are as follows: $|T| = 64 ,\text{max}(|\mathcal{C}_n|) = 21, \text{max}(|T^{\prime}|) = 4, |M^{\prime}| = 100$.

\begin{table}[t]
 \caption{Values of constants (performance evaluation)}
 \label{tab:para_performance}
 \centering
 \begin{tabular}{c|c||c|c}
   \bhline{0.7pt}
   \textbf{Parameter} & \textbf{Value} & \textbf{Parameter} & \textbf{value}\\
   \hline
   $A$ & $640$ & $K$ & $4$\\
   $B$ & $3200$ & $M^{\prime}$ & $100$\\
   $L$ & $16$ & $\distCloud$ & $5$ \\
   \bhline{0.7pt}
 \end{tabular}
\end{table}

Next, we conduct $|C_n|$ times notification deliveries for each topic $t_n \in \mathcal{T}$ , where the set of publishers $P_i$ is $c_m$ and the set of $S_i$ as other clients in $C_n$ for each $c_m \in C_n$. Thus, message publication and notification delivery to subscribers is conducted $I = \sum_{t_n \in T} |C_n|$ times in total. 

We evaluate the notification delivery delay by the metric $Y$ defined as follows:
\begin{eqnarray}
    Y &=& \dfrac{1}{I}\sum_{i=1}^{I} D_i \nonumber \\
     &=& \dfrac{1}{I}\sum_{i=1}^{I}\frac{1}{|S_i|}\sum_{c_m \in \mathcal{P}_i}\sum_{c_{m^{\prime}} \in \mathcal{S}_i} \Biggl[\sum_{l=1}^L \distCS{m}{l} x_{m,l}
  + \biggl[\sum_{l=1}^L 2\distCloud (1- \rl_l) x_{m, l} 
  +\sum_{l=1}^L \sum_{l^{\prime}=1}^L \rl_l \distSS{l}{l^{\prime}} x_{m, l}x_{m^{\prime} l^{\prime}}\biggr]
  + \sum_{l^{\prime}=1}^L \distCS{m^{\prime}}{l^{\prime}}x_{m^{\prime} l^{\prime}}\Biggr],
\end{eqnarray}
which is the average delay of the formulated cost shown in Equation~\ref{equ:objective}. Moreover, to observe a bottlenecks that cause delay, we define metrics $Y_1$ and $Y_2$:
\begin{eqnarray}
  Y_1 &=& \dfrac{1}{I}\sum_{i=1}^{I} \dfrac{1}{I}\frac{1}{|S_i|}\sum_{c_m \in \mathcal{P}_i}\sum_{c_{m^{\prime}} \in \mathcal{S}_i} \Biggl[\sum_{l=1}^L \distCS{m}{l} x_{m,l}
+ \sum_{l^{\prime}=1}^L \distCS{m^{\prime}}{l^{\prime}}x_{m^{\prime} l^{\prime}}\Biggr],\\
  Y_2 &=& \dfrac{1}{I}\sum_{i=1}^{I} \dfrac{1}{I}\frac{1}{|S_i|}\sum_{c_m \in \mathcal{P}_i}\sum_{c_{m^{\prime}} \in \mathcal{S}_i} 
\Biggl[\sum_{l=1}^L 2\distCloud (1- \rl_l) x_{m, l} 
+\sum_{l=1}^L \sum_{l^{\prime}=1}^L \rl_l \distSS{l}{l^{\prime}} x_{m, l}x_{m^{\prime} l^{\prime}}\Biggr].
\end{eqnarray}
$Y_1$ is the delay caused by distance, while $Y_2$ is the delay caused by resource consumption. In this simulation, we represent $d^{\text{H}}_{m, l}$ by 
\begin{equation}
  d^{\text{H}}_{m, l} = \gamma g_{m, l},
\end{equation}
where $\gamma$ is delay per a kilometer $g_{m,l}$ is the geographical distance between a client $c_m$ and her/his home server $s_l$. In this simulation we set $\gamma$ by $\gamma = 0.1$~ms.  

\begin{figure}[t]
  \begin{center}
      \includegraphics[width=18cm]{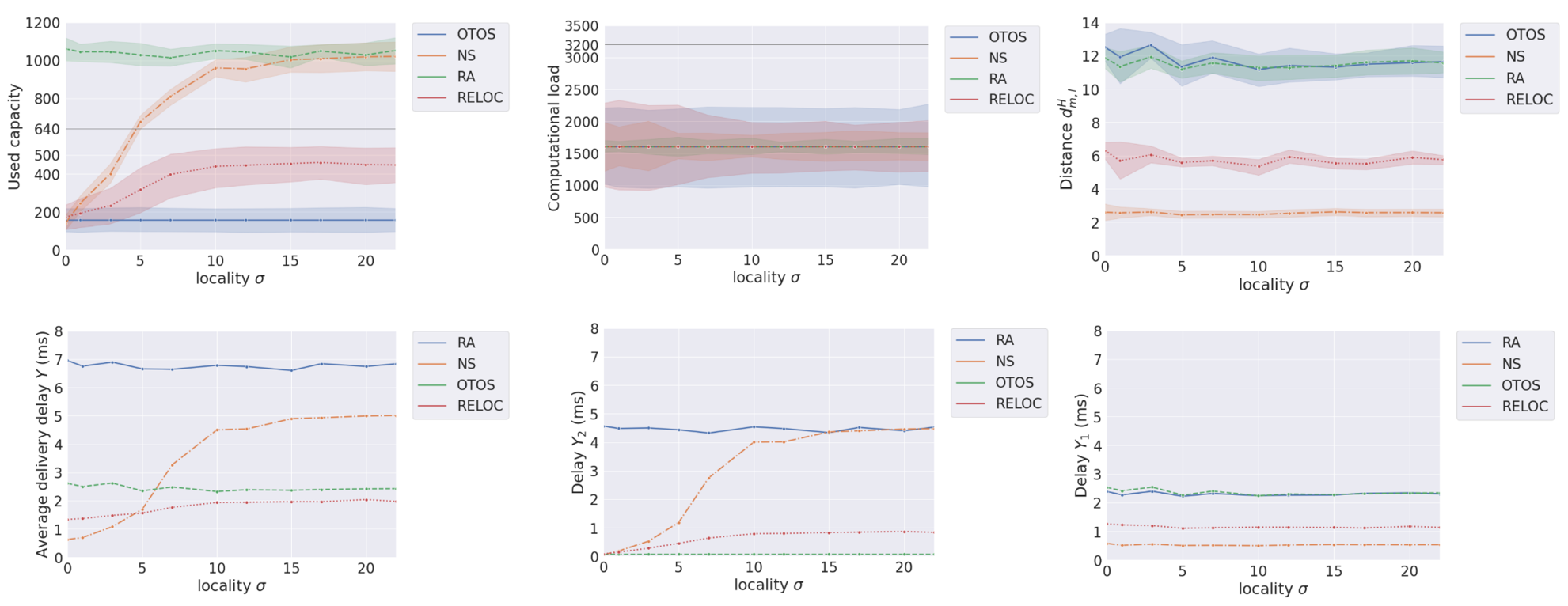}
    \caption{Comparing storage capacity usage, computational load, and distance between clients and their home server as well as average delivery delay $Y$, $Y_1$ and $Y_2$ of RA, NS, OTOS, and RELOC against locality of clients $\sigma$ ($p = 2$, $K = 4$).}
    \label{fig:result_locality}
  \end{center}
\end{figure}

FIGURE~\ref{fig:result_locality} shows storage capacity usage, computational load, and distance between clients and their home server as well as average delivery delay $Y$, $Y_1$ and $Y_2$ against locality of clients $\sigma$. All graph represents the average, and the band shown in the above three graph illustrates the $95\%$ confidential interval. 
From FIGURE~\ref{fig:result_locality}, we can observe that RELOC has the smallest delay to deliver notifications to clients if $\sigma \ge 5$. NS gives the smallest distance $d^{\text{H}}_{m,l}$, but NS gives delay $Y_2$ as the value of $\sigma$ increases. When there is sufficient capacity, NS method may be the best method. OTOS gives the smallest storage capacity consumption, but it gives more delay $Y_1$ than RELOC constantly. The black line in the graphs that show used capacity and computational load is $A$ and $B$ respectively. When applying RELOC, the storage capacity is not over $A$ even when locality $\sigma$ increases. This implies that the total storage consumption is reduced by \textit{relation conscious} step and storage capacity consumption is adjusted among edge servers by transferring message processors to servers which are not heavily used by \textit{cooperation} step. 

\begin{figure}[t]
  \begin{center}
      \includegraphics[width=18cm]{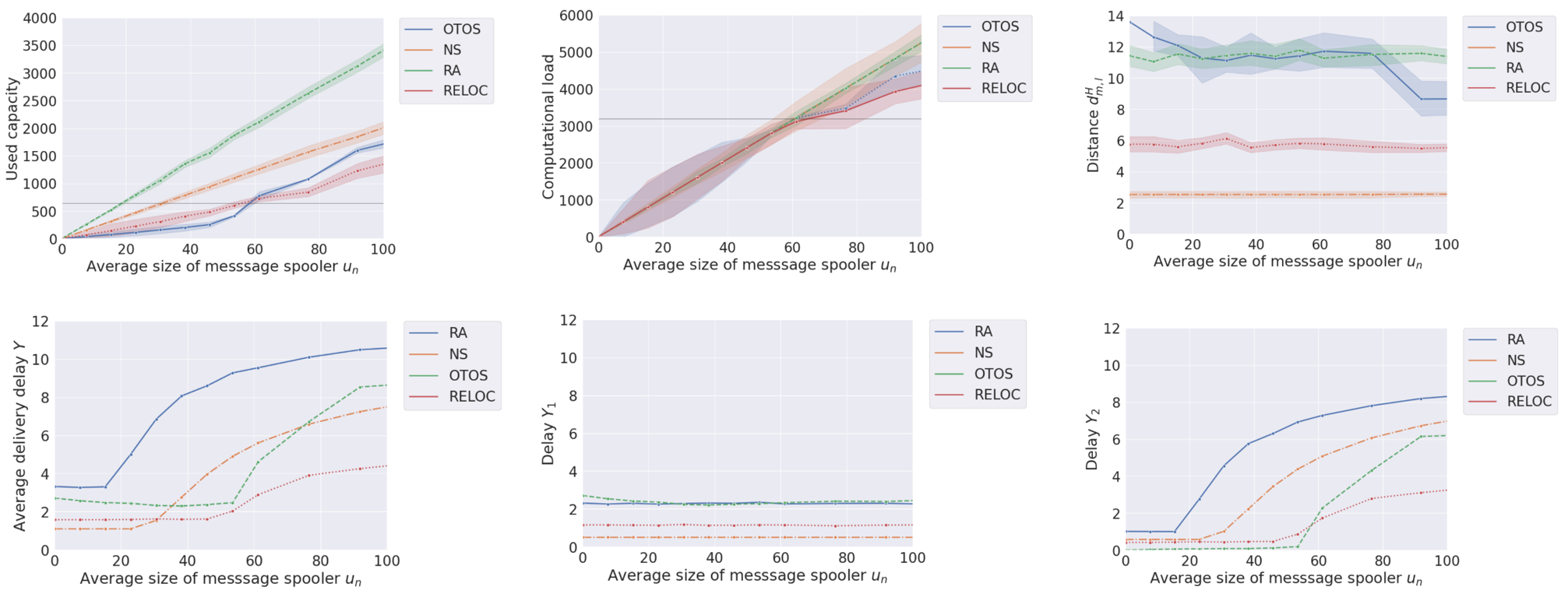}
    \caption{Comparing storage capacity usage, computational load, and distance between clients and their home server as well as average delivery delay $Y$, $Y_1$ and $Y_2$ of RA, NS, OTOS, and RELOC against average size of spoolers ($\sigma = 5$, $K = 4$)}
    \label{fig:result_spooler}
  \end{center}
\end{figure}

FIGURE~\ref{fig:result_spooler} shows storage capacity usage, computational load, and distance between clients and their home server as well as average delivery delay $Y$, $Y_1$ and $Y_2$ against the average size of message spooler for each topic $|u_n|$.
From FIGURE~\ref{fig:result_spooler}, we can observe that RELOC has the smallest delay to deliver notifications to clients. Even when used capacity and computational load is over the constraint $A$ and $B$ after \textit{relation conscious} step, the increase of average delivery delay $Y$ is small by virtue of \textit{cooperation} step.

\begin{figure}[t]
  \begin{center}
      \includegraphics[width=18cm]{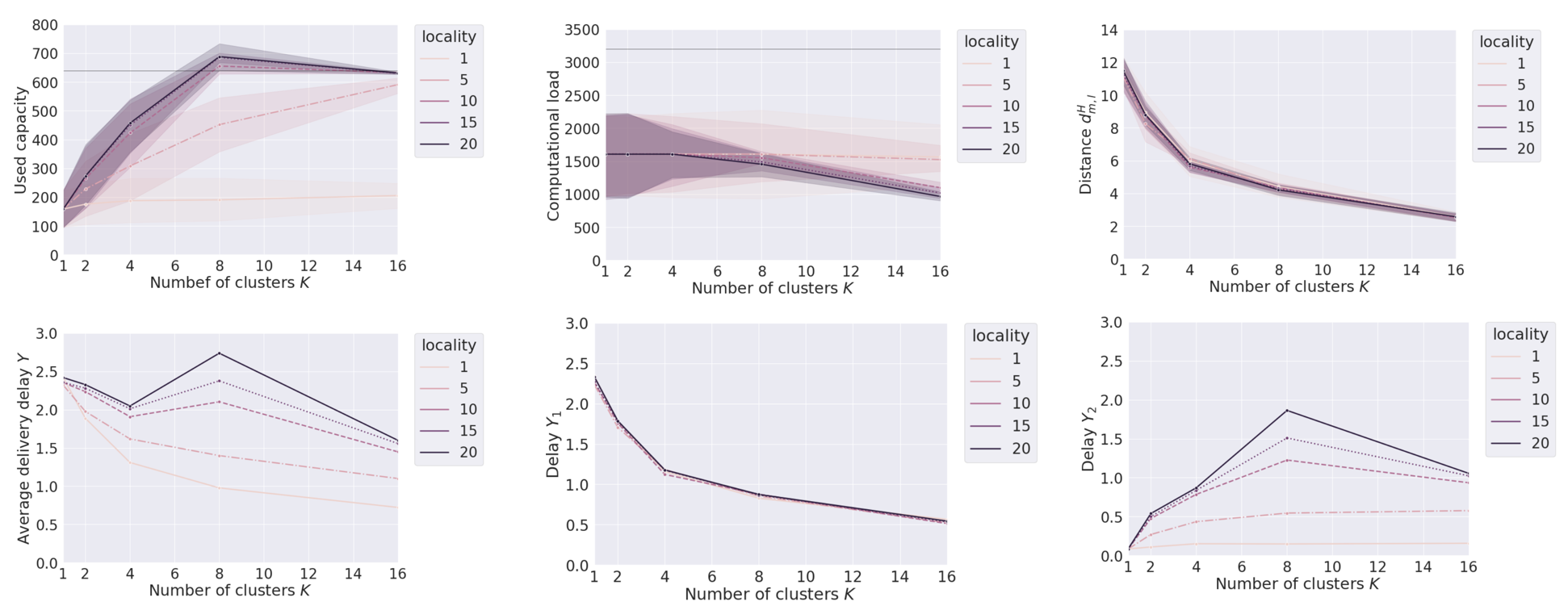}
    \caption{Comparing storage capacity usage, computational load, and distance between clients and their home server as well as average delivery delay $Y$, $Y_1$ and $Y_2$ of RELOC against the number of clusters $K$ ($p = 2$, method: RELOC)}
    \label{fig:result_cluster}
  \end{center}
\end{figure}

Finally, we investigate the adequate number of clusters $K$. FIGURE~\ref{fig:result_cluster} shows storage capacity usage, computational load, and distance between clients and their home server as well as average delivery delay $Y$, $Y_1$ and $Y_2$ against the number of clusters $K$. 
From FIGURE~\ref{fig:result_cluster}, when locality $\sigma = 10,15,20$, though the average delivery delay fluctuates depending on the number of clusters $K$, the average delivery delay does not vary so much. On the other hand, when locality is small, that is $\sigma = 1, 5$, the delay $Y$ is minimized by increasing the value of $K$.

\section{Conclusion}
In this paper, we presented a publish-process-subscribe system that allows messages published to a topic to be immediately processed by distributed message processors on edge servers. We formulated the delay caused by excessive edge server resource to find the optimized allocation of topics on edge servers. Numerical experiments show that our heuristics constructed with an analysis of the formulated optimization problem give an efficient use of edge server resources and reduce the message and notification delivery delay.

Future work involves comparing results by the optimal solution and heuristic solution, considering how to determine $K$ adequately in Algorithm~\ref{alg:reloc} exploiting several features of application users such as locality and mobility, and investigating time complexity and execution frequency to confirm that RELOC can be deployed to real environments.

\section*{Acknowledgement}
We would like to thank Patrick Finnerty for his precious comments. This work was supported by JSPS KAKENHI Grant Number JP18H03232, JP20K11841. This work is partly carried out on StarBED which is provided by National Institute of Information and Communications Technology (NICT).

\bibliography{reference}


\end{document}